\documentclass[final,twocolumn,sort&compress]{elsarticle}
\usepackage[utf8]{inputenc}
\usepackage[tbtags]{amsmath}
\usepackage{amssymb,amsthm,amsfonts,relsize,layout,indentfirst,graphicx,float,wrapfig,color,tabls}
\usepackage{boxedminipage}
\usepackage{lineno}
\usepackage[font={bf,small},textfont={it,md,small}]{subfig}
 \linenumbers
\begin{document}
%
%
	\title{Development of a Bayesian method for the analysis of inertial confinement fusion experiments on the NIF}
	\tnotetext[t1]{This work performed under the auspices of the U.S. Department of Energy by Lawrence Livermore National Laboratory under Contract DE-AC52-07NA27344. LLNL-JRNL-614352}
	\author[llnl]{J.A.~Gaffney\corref{cor1}}
	\ead{gaffney3@llnl.gov}
	\author[llnl]{D.~Clark}				
	\author[llnl]{V.~Sonnad}
	\author[llnl]{S.B.~Libby}
	\cortext[cor1]{Corresponding author}
	\address[llnl]{Lawrence Livermore National Laboratory, 7000 East Ave, Livermore, CA 94550}
%
%
%
%
	\begin{abstract}
	The complex nature of inertial confinement fusion (ICF) experiments results in a very large number of experimental parameters that are only known with limited reliability. These parameters, combined with the myriad physical models that govern target evolution, make the reliable extraction of physics from experimental campaigns very difficult. We develop an inference method that allows all important experimental parameters, and previous knowledge, to be taken into account when investigating underlying microphysics models. The result is framed as a modified $\chi^{2}$ analysis which is easy to implement in existing analyses, and quite portable. We present a first application to a recent convergent ablator experiment performed at the NIF, and investigate the effect of variations in all physical dimensions of the target (very difficult to do using other methods). We show that for well characterised targets in which dimensions vary at the 0.5\% level there is little effect, but 3\% variations change the results of inferences dramatically. Our Bayesian method allows particular inference results to be associated with prior errors in microphysics models; in our example, tuning the carbon opacity to match experimental data (i.e., ignoring prior knowledge) is equivalent to an assumed prior error of 400\% in the tabop opacity tables. This large error is unreasonable, underlining the importance of including prior knowledge in the analysis of these experiments.
	\end{abstract}
%
%
	\maketitle
%
%
\section{Introduction}
The design of experimental schemes to reach thermonuclear ignition and burn in laser driven targets involves complex models that incorporate many physical effects \cite{atzeni}. The radiation-hydrodynamic simulations used to predict the evolution of fusion capsules \cite{castor} therefore contain a huge number of physical parameters which play an important role. The resulting laser targets are correspondingly complex, with a large number of design parameters. In a typical inertial confinement fusion (ICF) experiment performed at large laser facilities such as the national ignition facility (NIF) \cite{nif}, there are many tens of variables that play an important role in determining target evolution \cite{haan11}. This poses a difficult problem for data analysis since these parameters should not be neglected but are too numerous to treat directly using the standard methods of, for example, particle physics where Monte Carlo sampling of noise sources is often used \cite{lyons08}. In this paper we develop a method that allows all important variables to be included, along with prior work on microphysics models, in a consistent and efficient analysis. The approach has been designed to couple with existing radiation-hydrodynamics simulation codes without modification; in fact simulations are treated as a `black box’ making the method applicable to a large class of difficult data analysis problems. This approach also allows us to avoid the complex fitting functions used in other approaches \cite{kennedy01, mcclarren11}, which are unlikely to capture the complex behavior of ICF experiments close to ignition (and are unsuitable for such large problems in any case).\par
The data analysis approach that we describe is particularly important when considering the results of the recent national ignition campaign (NIC). Throughout the campaign, post-shot simulations failed to match the observed data; the implication is that simulations, or their underlying microphysics models, are inaccurate. Determining which of the models should be investigated, and producing a consistent picture of the implied error, is a difficult task and forms a major motivation for this work.\par
In fact, modifications to various physical parameters, even unrealistically large modifications, often cannot produce a match to NIC data. In this situation the neglect of important variables and prior knowledge has a dramatic effect on the results of inference (even if they are purely sources of noise). Our method allows these effects to be included with almost no computational overhead. We demonstrate this by presenting an analysis of a single NIC convergent ablator (conA) shot \cite{hicks10,hicks12}, N110625. We find that variations in the dimensions of the target can have a dramatic effect on the inferred drive and carbon opacity, although this is mitigated by thorough metrology of the target. Our method also allows prior knowledge to be included and in the case of ICF we find that this is an extremely important factor. We use it to investigate the implied error in microphysics models associated with neglecting this prior work. The inclusion of this prior knowledge is an important strength of the Bayesian method, as it provides context for observed data and therefore allows meaningful information to be inferred even from a single experimental result. The total information from a set of experiments can be viewed as a series of such single-shot inferences, allowing the analysis performed here to be generalised to full experimental campaigns very easily.\par
The approach is to treat the output of the simulation code as probabilistic, and to apply standard methods of Bayesian analysis \cite{sivia}. The probabilistic nature of simulations is due to variations in the myriad important variables (or `nuisance parameters’). We derive a semi-analytic expression in which the dependence on interesting physics is retained but all other variables are represented by an analytic information loss. The result is framed as a modified $\chi^{2}$ analysis which is easy to implement, portable, and allows all available data to be included in a single analysis.\par
We begin by elaborating on the challenges we have already introduced. We then develop our inference approach in section \ref{sec:probabilistic_output}, and discuss methods for its application in section \ref{sec:implementation}. Finally, the importance of including all important variables and prior knowledge is demonstrated with an example application to a single NIC shot, N110625.\par
%
%
\section{Challenges for data analysis from ICF experiments} \label{sec:challenges}
In current analyses, particular data (chosen largely through experience) are preferentially matched by varying inputs that are considered to be unreliable, such as X ray drive \cite{robey12,gu12}. This approach has been very useful in testing the consistency between simulations and experiment, however it is potentially sensitive to noise and gives little information about the physical origin of inconsistencies. Increasing the number of inferred parameters is essential to gaining more information about underlying physics models.\par
Radiation-hydrodynamic simulations represent a nonlinear map from the space of physical models that we wish to investigate to the data that are collected in an experiment. The nature of the simulations often means that they are not amenable to adjoint differentiation \cite{hanson98}, are discontinuous, and may be noisy; these complex features can make standard methods of searching the space of physical parameters quite unreliable. This limits the number of parameters that can be reliably inferred. The nuisance parameters included by our method result in a smoothing of the code output, allowing the use of advanced methods and an increase in the number of physical parameters that can be investigated.\par
We have already described the difficulties associated with the large numbers of target parameters involved in ICF experiments. Although many of these are constrained by manufacturing precision and target metrology, it has already been seen that their large number can have an important impact on the output of simulation codes \cite{haan11}. There will be a corresponding effect on inference results, and we aim to investigate this.\par
The physical parameters we aim to infer often refer to microscopic physics (for example opacities or equations of state) that are understood using other, separate, computer simulations. These simulations are highly complex and have been investigated both theoretically and experimentally for many decades; the expected systematic error bars on their outputs are therefore quite small. This error bar plays an important role in data analysis by ensuring that the results are physically reasonable, and this motivates our Bayesian approach.\par
%
%
\section{Probabilistic output from a deterministic simulation code – the importance of nuisance parameters} \label{sec:probabilistic_output}
%
%
The fundamental problem is to develop a method of exploring the huge space of parameters that can affect the outcome of a simulation. As discussed, in the case of ICF data there is no point in this space for which all data are correctly simulated. In general there may be a set of points that give comparable agreement. The best fit is found by defining a figure of merit that takes into account the difference between observed and simulated values of all data points, as well as the difference between simulation parameters and the expected physical reality. In this section we outline a figure of merit that is based on the Bayesian posterior probability of a point in phase space (the \emph{maximum a posteriori}, or MAP, solution \cite{sivia}), and use an analytic prior-predictive approach to reduce the phase space to manageable size.\par
We begin by splitting the set of all parameters into two;
\begin{itemize}
	\item `Interesting Parameters' $\theta$ - Physically significant parameters that we aim to infer from experiment data. For example material equation of state, opacities, conductivities, ...,
	\item `Nuisance Parameters' $\eta$ – Other parameters that have an effect on simulations but are not of direct physical significance. These are usually known with good precision, for example target dimensions, laser powers, ...,
\end{itemize}
\par
In our model, inference is performed on the interesting parameters only. Bayes' theorem allows us to write down the probability distribution of the interesting parameters once the experiment has been performed (the \emph{posterior}), in terms of the probability distribution before the experiment (the \emph{prior}) and the probability of the experimental data (the \emph{likelihood}). Bayes' theorem is
\begin{align}
	P(\theta| d_{exp}) &= \frac{P(d_{exp}|\theta)P(\theta)}{P(d_{exp})} \notag\\
	&=\frac{ \iint d\eta~ dd_{m} ~ P(d_{exp},d_{m},\theta,\eta)}{P(d_{exp})} \notag
	~{\rm ,}
\end{align}
where $d_{exp}$ is the vector of experimental data and we have introduced the code output $d_{m}$ and the nuisance parameters as marginalised variables. This allows us to introduce the known measurement error and prior distributions of the nuisance parameters later. Such an approach is equivalent to assuming that experimental data are the simulation results plus a randomly distributed error, as is done in other approaches \cite{kennedy01,higdon08}. Writing $P(d_{exp},d_{m},\theta,\eta)=P(d_{exp}|d_{m},\theta,\eta)P(d_{m}|\theta,\eta)P(\theta,\eta)$ and introducing the deterministic nature of the simulation code,
\begin{equation*}
	P(d_{m}|\theta,\eta)=\delta(d_{m}-d_{m}(\theta,\eta)) ~{\rm ,}
\end{equation*}
the integration over $d_{m}$ can be performed trivially. The result is
\begin{align}
	P(\theta| d_{exp}) &= \frac{P(\theta)}{P(d_{exp})} \int d\eta ~ P(d_{exp}|d_{m}(\theta,\eta))P(\eta) \notag\\
	&\equiv\frac{P(\theta)}{P(d_{exp})} P(d_{exp}|\theta)
	~{\rm ,}
	\label{eq:bayes_theorem}
\end{align}
The likelihood $P(d_{exp}|d_{m}(\theta,\eta))$ implicitly contains the experimental error distribution and the code output as a function of all parameters $d_{m}(\theta,\eta)$. The two components of the prior distribution $P(\theta,\eta)\equiv P(\theta)P(\eta)$ describe the expected distributions of nuisance and interesting parameters before the experiment has been performed; these are determined by the experimental design, target manufacturing tolerances, previous experimental results and expert opinion.\par
Equation \eqref{eq:bayes_theorem} describes the relationship between the probability distributions of the interesting parameters before and after an experiment. The details of the relationship are approximated by the simulation code, and contained in the likelihood function. Data analysis, then, is based on the evaluation of the integral in the definition of the likelihood. In its general form this involves the integration of simulation output over the entire nuisance parameter space; it is common to evaluate this using a Monte-Carlo sampling of the integrand (see, for example, \cite{roe07}). For our application, where even a conservative set of nuisance parameters results in a $\sim20$ dimensional integral, this is prohibitively expensive. Even if the radiation-hydrodynamics can be modelled by some fast surrogate model (as a Gaussian Process or through other techniques \cite{sacks89,kennedy01,mcclarren11}), which itself is very difficult given the size of the space we must consider, the integral is still too expensive. We instead evaluate the integral by assuming a linear response to nuisance parameters,
\begin{equation}
	d_{m}(\theta, \eta) = d_{m}(\theta, \eta=\eta_{0})  + A(\eta_{0}-\eta) ~{\rm .}
	\label{eq:linear_response}
\end{equation}
In the above $\eta_{0}$ is the nominal value of the nuisance parameters, typically zero. The linear response matrix $A$ can be populated using a small number of simulations; once this has been done, the matrix $A$ is entirely portable and may be used in all subsequent analyses of this type without further calculation.\par
%
%
The case of linear response, equation \eqref{eq:linear_response}, is very useful as it allows an analytic treatment of the nuisance parameters. Assuming that the experimental measurement errors and nuisance parameter variations are described by uncorrelated normal distributions with correlation matrices $\Lambda_{exp}$ and $\Lambda_{\eta}$,
\begin{equation}
	\begin{split}
	P(d_{exp} \vert \theta) = \int d\eta ~ \biggl\{ &
		\frac{ e^{-(d_{exp}-d_{m}(\theta,\eta))^{T}\Lambda_{exp}^{-1}(d_{exp}-d_{m}(\theta,\eta))}}{\sqrt{(2\pi)^{n_{exp}}|\Lambda_{exp}|}} \times \\
	& \times \frac{ e^{-(\eta-\eta_{0})^{T}\Lambda_{\eta}^{-1}(\eta-\eta_{0})}}{\sqrt{(2\pi)^{n_{\eta}}|\Lambda_{\eta}|}}
	\biggr\}
	~{\rm ,} \notag
	\end{split}
\end{equation}
the result is
\begin{equation}
	P(d_{exp} \vert \theta) = \frac{e^{-(d_{exp}-d_{m}(\theta))^{T} \left[ \Lambda_{exp}^{-1}-\beta^{T}\beta\right] (d_{exp}-d_{m}(\theta))}}{\sqrt{(2\pi)^{n_{exp}}|\Lambda_{exp}||\Lambda_{\eta}||\alpha^{T}\alpha|}} 
	~{\rm .}
	\label{eq:likelihood_marginal}
\end{equation}
In the above, $d_{m}(\theta)\equiv d_{m}(\theta,\eta_{0})$ is the simulation result for nominal nuisance parameters and the matrices $\alpha$ and $\beta$ satisfy the equations
\begin{align}
	\alpha^{T}\alpha &=A^{T}\Lambda_{exp}^{-1}A+\Lambda_{\eta}^{-1} \notag \\
	\beta^{T}\alpha &= \Lambda_{exp}^{-1}A \notag
	~{\rm .}
\end{align}
These expressions are the multivariate generalisation of the usual quadrature error propagation formula; it should be noted that even if nuisance parameters and experimental errors are independent to begin with (i.e., if $\Lambda_{exp}$ and $\Lambda_{\eta}$ are diagonal), the response of the simulations means that the likelihood can become strongly correlated. These potentially strong correlations arise due to the deterministic nature of the simulation code and play a very important role in the inference procedure described in the next section.\par
\section{Inference of interesting parameters from experimental data} \label{sec:implementation}
The results of the previous section allow the efficient calculation of the likelihood as a function of interesting parameters, without neglecting other important variables or prior knowledge. As discussed in section \ref{sec:challenges}, this can be expected to give a significant improvement in data analysis results. The marginalisation of nuisance parameters represents an averaging that smooths the response of simulations, making them more well-behaved. This allows us to use standard numerical techniques.\par
The best fit to data, taking into account nuisance parameters and prior knowledge, is given by the parameters that maximise the posterior probability $P(\theta|d_{exp})$ (see equation \eqref{eq:bayes_theorem}). It is convenient to minimise the information, $I(\theta|d_{exp}) = -LogP(\theta|d_{exp})$, which using equations \eqref{eq:bayes_theorem} and \eqref{eq:likelihood_marginal} is
\begin{equation}
	\begin{split}
		I(\theta|d_{exp})=& \sum_{i} \frac{(d_{exp,i}-d_{m}(\theta)_{i})^{2}}{\sigma_{exp,i}^{2}}  \\
			 & - (d_{exp}-d_{m}(\theta))^{T} \beta^{T}\beta (d_{exp}-d_{m}(\theta)) \\
			 & +\frac{1}{2}{\rm ln}\left(|\Lambda_{\eta}||\alpha^{T}\alpha|\right) - {\rm ln}P(\theta)
	\end{split}
	~{\rm .}
	\label{eq:information_likelihood_marginal}
\end{equation}
The above equation has the form of a modified $\chi^{2}$ function, and is derived by assuming that $\Lambda_{exp}$ is diagonal. Note that the dependence of the first term on $\theta$ through the simulation $d_{m}(\theta)$ means that even in the absence of nuisance parameters and prior knowledge the likelihood is non-normal. Equation \eqref{eq:information_likelihood_marginal} can be interpreted as an information processing rule \cite{zellner88}; the first 3 terms on the right hand side are the information gained from the experiment, and the final term is the information about the interesting parameters before the experiment was performed. In that sense it is clear that the positive definite matrix $\beta^{T}\beta$ represents a loss of information due to nuisance parameters. As mentioned, once $\beta^{T}\beta$ has been computed the evaluation of the modified $\chi^{2}$ only requires a single simulation.\par
In an actual inference problem we are interested in the values of $\theta$ that give the best fit to the experimental data. This requires the numerical minimisation of equation \eqref{eq:information_likelihood_marginal}. The well behaved nature of the marginalised likelihood allows us to use standard methods; two common approaches are
\begin{itemize}
	\item Markov Chain Monte Carlo (MCMC) – This approach gives an approximation to the entire posterior information \cite{green95}. This is extremely useful to the evaluation of error bars on inferred parameters. The trade-off is that these methods require extensive ‘burn in’ periods and are difficult to run in parallel. In our applications where a single simulation represents a significant computational overhead, this is a major disadvantage;
	\item Genetic Algorithm (GA) – This method uses ideas taken from genetics to efficiently find the minimum of a function. It is very easy to run in parallel and so is well suited to our application. The final result is the position of the minimum only and so some approximation is required to calculate error bars \cite{sivia},
\end{itemize}
\par
In the following section we present an example application of the method. For simplicity the parameter space is small allowing the likelihood and posterior to be explored directly. The advanced techniques discussed above are therefore not needed. In a forthcoming paper we consider more complex cases for which we develop a genetic algorithm that is optimised for the sparse datasets encountered in ICF research.\par
%
%
\section{Application to NIF convergent ablator experimental data}
In order to demonstrate the application to an actual inference problem, we now consider experimental data taken on a NIC convergent ablator (conA) experiment \cite{hicks10, hicks12}. In this design, an ICF capsule is imploded and backlit by emission from a nearby high $Z$ plasma. This allows time- and space- resolved measurement of the plasma density during the implosion. This is analysed to give time resolved measurement of fuel shell position, velocity, and line density. Simple models show that these quantities are sensitive to the details of the X ray drive from the hohlraum, and to radiation transport in the capsule ablator \cite{atzeni}.\par
In these experiments the measured implosion velocities are consistently lower than simulated predictions, possibly suggesting a reduced X ray drive. Absorption of the drive X rays by carbon in the ablator plastic also plays an important role, and simulations show that an increased carbon opacity can improve agreement \cite{clark-pres11}. These parameters can be used to tune simulations to agree with experiment, however such an approach runs the risk of destroying the predictive capabilities of codes when run far from existing experimental data (a common occurrence in all areas of HEDP, and a possible problem when designing improved ICF targets). We aim to analyse the significance of these experimental results with respect to the drive and carbon opacity by inferring the values of modifiers to those quantities in the presence of many nuisance parameters and of prior knowledge. The prior distributions that we place on these multipliers are interpreted as the uncertainties in the off-line calculations of opacity and drive.\par
The inference is based on the HYDRA radiation-hydrodynamics code \cite{hydra}. The parameters of interest are the values of two dimensionless multipliers; one is applied to the X ray drive spectrum (at all times and photon energies) and the other is applied to the carbon opacity (all temperatures, densities and photon energies). We take into account 29 nuisance parameters, allowing all capsule dimensions, material densities and material compositions \cite{haan11} to vary. These parameters are allowed to vary according to two distributions with standard deviations of $0.5\%$ and $3\%$ respectively. These represent well constrained target parameters (many NIC capsule dimensions are known to better than $0.5\%$), and ones with more uncertainty (demonstrating the potential effect of nuisance parameters on inferred physics). For this relatively small inference it is feasible to generate a set of simulations that span the 2D parameter space. For each point, defined by the multipliers $(\Delta_{drive}, \Delta_{C opac})$, we run HYDRA and extract the implosion velocity, ablator mass fraction, and time at which the implosion reaches a radius of $310\mu$m. These quantities are compared to experimental values taken from radiography \cite{hicks}.\par
\begin{figure}
	\centering
	\subfloat[No nuisance parameters]{
		\includegraphics[scale=0.25]{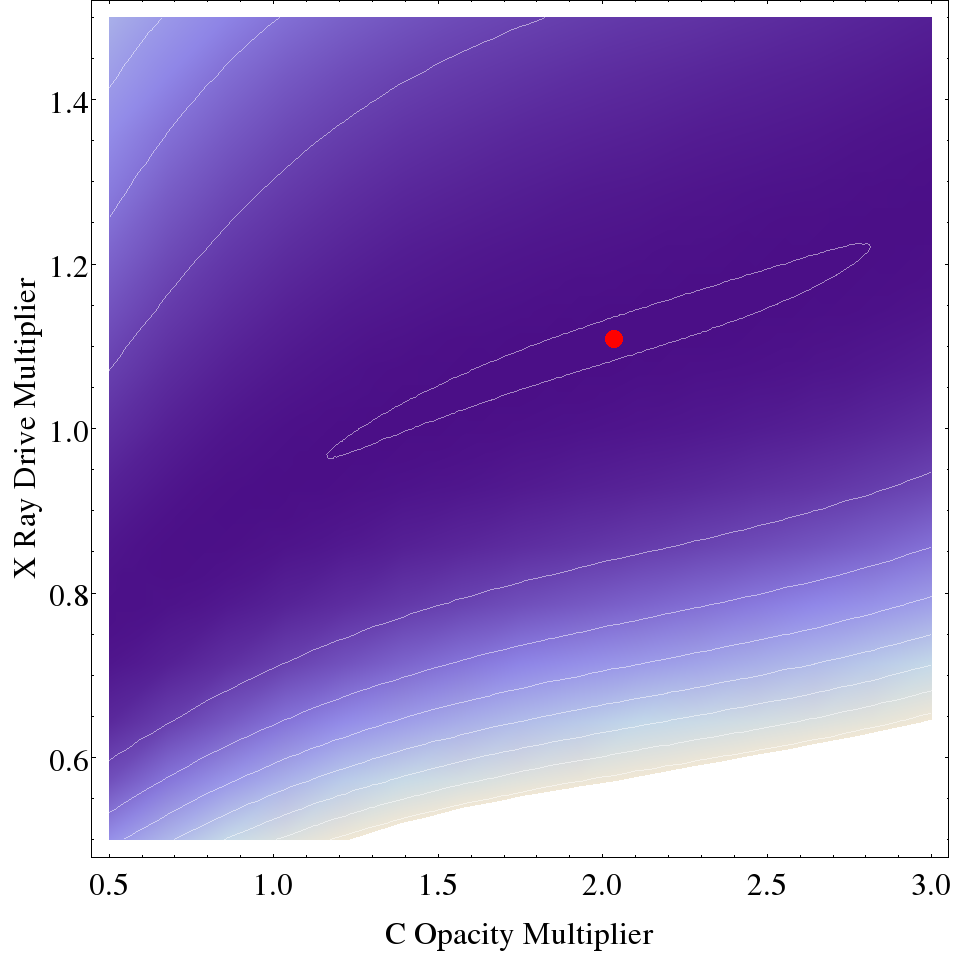}
	}\\
	\subfloat[Linear model for nuisance parameters with 3\% variations]{
		\includegraphics[scale=0.25]{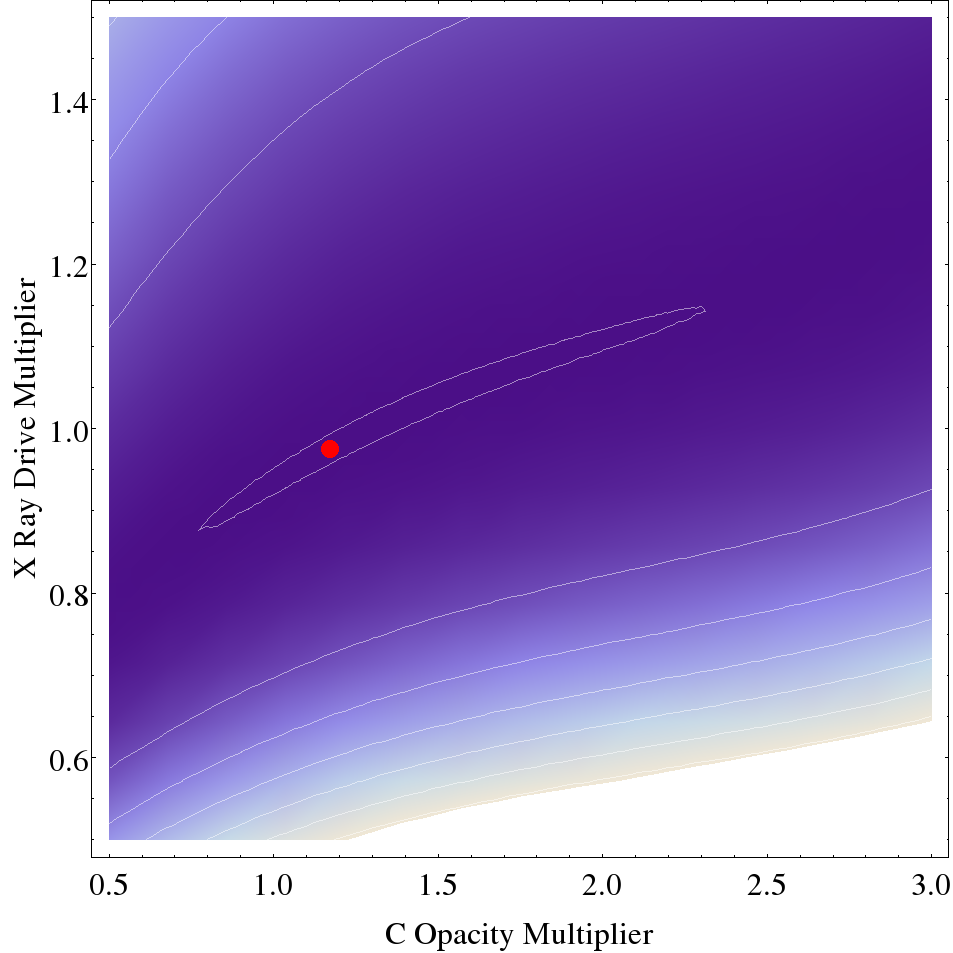}
	}
	\caption{Information in the likelihood for multipliers placed on the carbon opacity and X ray drive for a NIC conA experiment. Panel (a) shows the result when no nuisance parameters are included, and (b) shows the effect of including target metrology as nuisance parameters.}
	\label{fig:likelyhoods}
\end{figure}
In figure \ref{fig:likelyhoods} we plot the information in the likelihood as a function of $\Delta_{drive}$ and $\Delta_{C opac}$, calculated using equation \eqref{eq:likelihood_marginal} with different values of the modification matrix $\beta^{T}\beta$. These plots represent our modified $\chi^{2}$ when the prior distribution $P(\theta)$ is neglected. In (a) nuisance parameters are neglected ($\beta^{T}\beta=0$), and in (b) the modification is calculated as described for all 29 nuisance parameters varying at the $3\%$ level. In the case with $0.5\%$ variations, nuisance parameters have a small effect and the likelihood is very similar to figure \ref{fig:likelyhoods}(a). The positions of the minima are marked with red points, making the effect of nuisance parameters clear. This shift in minimum is very important in the subsequent analysis.\par
To further quantify the differences we perform a set of inferences based on the calculated likelihoods. The specific choice of prior distribution is often a difficult issue since it can be a subjective choice that has a direct influence on inference results. For this reason we perform a range of inferences with prior distributions for $(\Delta_{drive}, \Delta_{C opac})$ of varying width. This allows us to take into account the dependence of the inference results on the prior, and place limits on the actual prior for various results. 
%
\par
We begin with a reasonable estimate of the uncertainties in opacity and drive models of 10\% and 20\% respectively. This defines our nominal prior as a normal distribution centered on $(1,1)$, with covariance matrix
\begin{equation}
 	\Lambda_{p} = \begin{pmatrix}
 		0.1^{2} & 0 \\
 		0 & 0.2^{2}
 		\end{pmatrix}
 	\label{eq:prior_covariance}
	~{\rm .}
\end{equation}
A set of inference results are found by scaling this covariance, thereby changing the assumed prior error in microphysics models (and the relative importance of prior and experimental information). For a very large scaling of \eqref{eq:prior_covariance}, the prior is flat and our analysis reproduces the maximum likelihood (ML) result; for a small scaling factor the prior tends to a $\delta$-function and the minimum of equation \eqref{eq:information_likelihood_marginal} is at $(\Delta_{drive}, \Delta_{C opac}) = (1,1)$ (their prior values). \par
\begin{figure}
	\centering
	\includegraphics[scale=0.25]{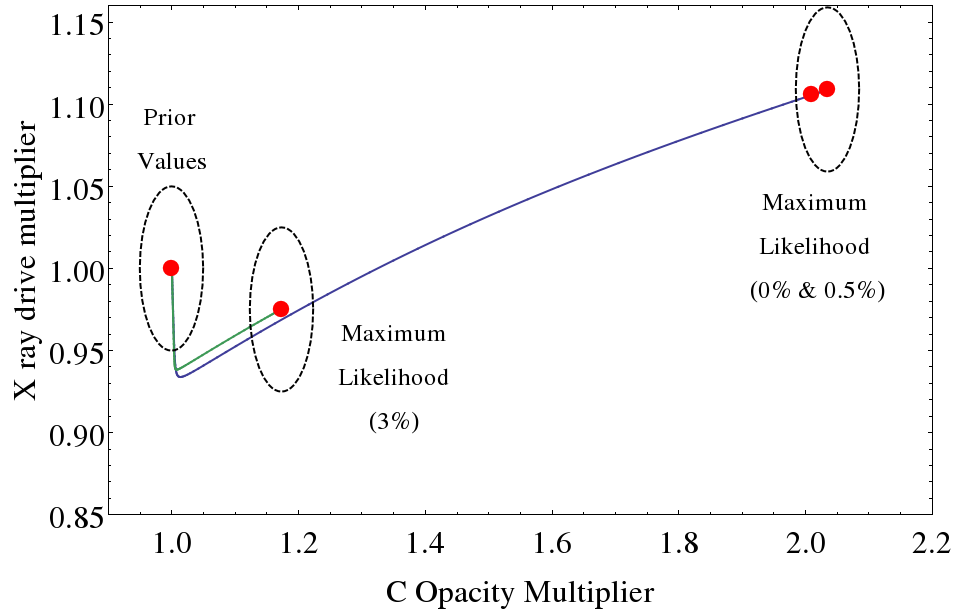}
	\caption{Trajectories of the best fit to experimental data from a NIC conA experiment, as the prior width is varied (see equation \eqref{eq:prior_covariance}). The blue line shows the result when nuisance parameters are ignored, or included at the $0.5\%$ level. The two red points at the right hand end represent the maxima of the likelihood for these two cases. The green line shows the case when nuisance parameters are included at the $3\%$ level. As the prior is scaled from a $\delta$-function, through our best estimated defined by \eqref{eq:prior_covariance}, to flat, the inferred results tracks from the prior results $(1,1)$ to the minimum of the likelihood functions plotted in figure \ref{fig:likelyhoods}. The figure also shows contours that define a change in multiplier of 5\% from each end point.}
	\label{fig:posterior_tracks}
\end{figure}
%
%
In figure \ref{fig:posterior_tracks} we plot the trajectories of the best fit as the prior covariance is scaled from small to large. The trajectory for calculations that neglect nuisance parameters, and that include them at the $0.5\%$ level, overlay each other and are shown in purple; note the slight shift in the ML result at the right hand end. The $3\%$ case is plotted in green. The shapes of the trajectories are determined by all the factors we have discussed so far, not least the shape of the likelihood (i.e., the effect of nuisance parameters). The left hand end of the trajectories corresponds to small prior error and reproduces the prior result. The right hand end of each line is the flat prior result; as we have already seen in figure \ref{fig:likelyhoods} the inclusion of nuisance parameters at the $3\%$ level has a very significant effect on the inferred values of our interesting parameters.\par
%
%
The wide difference between the start and end points of all trajectories in figure \ref{fig:posterior_tracks} clearly shows that the prior distribution has an extremely important role to play in our analysis. For our nominal prior, defined by the covariance \eqref{eq:prior_covariance}, we find that the prior is in fact more important than the details of the nuisance parameters regardless of their distribution widths, giving inference results that are almost the same; $(\Delta_{drive}, \Delta_{C opac}) = (1.03,0.94)$. The ML analysis, that neglects the prior, will then result in a significantly different result. This is true even for extremely broad priors; for our MAP analysis (which includes both prior and nuisance parameters) to reproduce the $0.5\%$ nuisance parameter ML result to within 5\% (shown by the dashed contours in figure \ref{fig:posterior_tracks}), the prior covariance must be scaled so that the prior errors in opacity and drive are more than 400\% and 800\% respectively. The simulations on which the opacity and drive are based can be expected to be much more accurate that this, giving further support to the importance of the prior.\par
%
%
%
\section{Discussion and Conclusions}
We have developed a Bayesian model for investigation of underlying physics using complex HED experiments. The model allows for the inclusion of complications arising in experiments by using an approximate description of so-called nuisance parameters, and of previous investigations through a Bayesian prior. The result is a modified $\chi^{2}$ function that can be easily incorporated into any analysis using standard methods. This approach allows complex simulations to be treated as black box transformations from physical models to experimental data and so is suitable for application in a wide range of physical applications.
The linear response model described is the basis of the usual `Normal Linear' model \cite{sivia}. However, unlike that model, the use of complex simulations to describe interesting parameters and the resultant correlations between nuisance parameters results in a non-normal posterior.\par
In the case of ICF experiments, the linear response approximation may not be sufficient. The difficult task of achieving thermonuclear ignition requires that target designs are highly optimised; a change in nuisance parameters in either direction is likely to produce a reduction in target performance. Such nonlinear behavior can be important, and is not captured by the current approach. Test calculations for a reduced problem, including quadratic response to nuisance parameters, suggests that these effects are significant in the analysis of ICF data. A major piece of further work is to develop an efficient way of including nonlinearity.\par
In the final sections of this paper we have applied our analysis to a single NIC experiment. We attempt to describe deficiencies in radiation transport physics through multipliers on two physical quantities, and infer the posterior values of these multipliers. This process is a common one in the analysis of NIF data, and is usually viewed as the tuning of simulations to allow more reliable target design. In this work we interpret the results of this process as a measure of the uncertainty in the underlying physical models, which are often applied in regimes where they are untested. Only by improvement of these models, motivated by the kind of data analysis described here, can a truly predictive simulation be developed.\par
The particular example given here is sufficient to demonstrate the importance of an integrated approach to data analysis, and provides compelling evidence that a straightforward fit to experimental data, ignoring prior knowledge, can give misleading results. For the very well characterised targets used at the NIF, certain dimensions are known to better than the $0.5\%$ accuracy we allow in this work, however other nuisance parameters (for example material densities) could vary over a larger range. We have demonstrated that these nuisance parameters may have an important effect; our method allows a complete description of the problem. Alongside the nuisance parameters that we have included in this demonstration, there are also many other simulation inputs which can be treated as nuisance parameters in the same way.\par
We demonstrated a novel method of analysing the importance of prior knowledge by referencing the possible conclusions from data to limits on prior distribution widths. The multipliers used here do not, however, provide an insight into specific problems in underlying physics; it is also true that these multipliers only describe the average modification to theory that is required. Any inferred physical modifier will lose its meaning when the simulations used in the inference have other unknown inaccuracies, and this is certainly the case in our first application. We begin addressing these problems in a forthcoming paper.\par
The work presented here represents the first steps to providing a clearer view of problems with physics models from experimental data, in cases where the experiments are very complex. Although we concentrate on ICF experiments here, nuisance parameters can be expected to be important in all HED experiments, in particular those where target plasmas are less well constrained. The portability of our method makes its application to other experiments very easy. The computational framework described also provides the opportunity for Bayesian experimental design \cite{chaloner95}, allowing future experiments to provide a significant measurement of difficult aspects of underlying physics \cite{fischer05}. The integrated approach that we propose may also facilitate discovery of new rules and phenomenology that govern the evolution of these complex systems.\par
%
%
%
	\refstepcounter{section}									
	\bibliographystyle{./phnf}								
	%

%
%
	\end{document}